\documentclass[12pt,preprint]{aastex}

\slugcomment{Accepted ApJL, May 2010}

\shorttitle{Non-Linear Schmidt Law}
\shortauthors{Madore}

\begin{document}


\title{Star Formation Timescales and the Schmidt Law}


\author{\bf Barry F. Madore}
\affil{Observatories of the Carnegie Institution for Science \\ 813
Santa Barbara St., Pasadena, CA ~~91101} \email{barry@obs.carnegiescience.edu}



\begin{abstract}
We offer a simple parameterization of the rate of star formation in
galaxies. In this new approach, we make explicit and decouple the
timescales associated (a) with disruptive effects the star formation
event itself, from (b) the timescales associated with the cloud
assembly and collapse mechanisms leading up to star formation.  The
star formation law in near-by galaxies, as measured on sub-kiloparsec
scales, has recently been shown by Bigiel et al. to be distinctly
non-linear in its dependence on total gas density. Our
parameterization of the spatially resolved Schmidt-Sanduleak relation
naturally accommodates that dependence.  The parameterized form of the
relation is $\rho_* \sim \epsilon\rho_g/(\tau_s + \rho_g^{-n})$, where
$\rho_g$ is the gas density, $\epsilon$ is the efficiency of
converting gas into stars, and $\rho_g^{-n}$ captures the physics of
cloud collapse. Accordingly at high gas densities quiescent star
formation is predicted to progress as $\rho_* \sim \rho_g$, while at
low gas densities $\rho_* \sim \rho_g^{1+n}$, as is now generally
observed.  A variable efficiency in locally converting gas into stars
as well as the unknown plane thickness variations from galaxy to
galaxy, and radially within a given galaxy, can readily account for
the empirical scatter in the observed (surface density rather than
volume density) relations, and also plausibly account for the noted
upturn in the relation at very high apparent projected column
densities.

\end{abstract}

\
\
\
\
\
\

\vfill\eject
\section{Introduction}

The Schmidt Law has become widely synonymous with any power-law
relation between a (local or even global) star formation rate
indicator (be it individual OB stars, HII regions, H$\alpha$
luminosity, UV luminosity, re-radiated 24$\mu$m radiation or the
strength of a [C IV] cooling line) and a corresponding measure of gas
density (originally HI and now more frequently H$_2$, or a summed
combination of the two)\footnote{For a comprehensive and tutorial
review see Leroy et al. (2008), Section 2 and especially Table 1.}.
Schmidt (1957, 1963) first proposed such a formalism after noting that
within the context of the Milky Way the gas scale height (the fuel for
star formation) was larger than the O-star scale height (the result of
star formation) suggesting a non-linear (plausibly a power-law)
causally connected relationship between the two.  Schmidt concluded
that the exponent connecting gas density to star formation had a value
of n $\sim$ 2. Subsequently, (with the exception of a solitary paper
by Guibery, Lequeux \& Viallefond 1978, again dealing with the Milky
Way star formation and gas scale heights) little interest in this
topic was visible for more than a decade.

The field became active again when some years later Sanduleak (1968)
offered a novel calibration of the Schmidt Law in an extragalactic
context.  Rather than working with scale heights he examined the
relationship between neutral hydrogen gas surface densities and the
projected surface densities of recently-formed and individually
resolved OB stars.  We refer hereafter to this spatially resolved
correlation of star formation tracers with projected gas surface
density as the Schmidt-Sanduleak Law, so as to clearly distinguish it
from the global relation linking the total star formation rate with
total gas content, often referred to as the Schmidt-Kennicutt Law. For
OB stars and neutral hydrogen gas in the SMC Sanduleak found
$\Sigma_{stars} \sim \Sigma^{1.84 \pm 0.14}_{gas}$.

Sanduleak's particular technique in his pioneering study was extended
to other galaxies and in the process also generalized to different
star-formation tracers.  For instance, Hartwick (1971) correlated HII
region surface densities with neutral hydrogen surface densities for
the Local Group galaxy M31, and derived an exponent $n = 3.50\pm0.12
$. Madore, van den Bergh \& Rogstad (1974) combined the methods of
Sanduleak and Hartwick and correlated both star counts and HII regions
with HI surface densities across the face of M33. They found
differences\footnote{And with the clarity of hindsight one can see a
flattening of the power-law relation at high gas densities in the M33
data both for the stars and for the HII regions (their Figures 3 and
4); however it was clearly too early in the game to be introducing
even more variables than had already been considered in this early
paper.}  between between the two tracers as well as differences as a
function of radius; the latter, being suggested by them, as being
plausibly due to plane thickness variations. They found exponents
ranging from $n = 0.57$ to $2.57$, depending on the tracer and the radial
region considered, with the outer regions showing a steeper
dependence. Tosa \& Hamajima (1975), Hamajima \& Tosa, (1975) followed
Hartwick's lead and also used HII region number densities and
correlated these with the available neutral hydrogen surface
densities. They found exponents ranging from $n = 1.5$ to $3.5$ in 7
galaxies (SMC, LMC, M31, M51, M101, NGC~2403 and NGC~6946). Their
analysis again suggested that the relation was a function of radius
where in the examples of M31 and M101 the exponent was again found to
be larger in the outer regions when compared to the inner disk.
More recently, Bigiel et al. (2008) examined the spatially resolved
star formation law in 18 nearby galaxies using both UV (GALEX) and IR
(Spitzer) data to characterize the on-going star formation, and
exploring both molecular, neutral and total hydrogen gas content as
control variables. They found that for total gas (surface) densities
in excess of 1M$_{\sun}$ pc$^{-2}$ the star formation rate scaled
linearly with total gas (i.e., $n = 1.0 \pm$0.2), but at lower gas
surface densities the power law steepens considerably and is widely
dispersed (see their composite Figures 10 and 15, and Figure 1 below). 

We now proceed to give a physically motivated, analytic expression
capturing these facts. It is hoped that with this formulation galaxy
evolution modellers will be able to exploit this simple parametric
means of predicting star formation rates based on local gas densities.

\section{Timescale Arguments}

Numerical simulations of star formation, in search of the underlying
parameterization of the causal relation between observed gas densities
and current star formation rates, were first presented by Madore
(1977). There it was pointed out that if the star formation rate was
dimensionally decomposed into its native sub-components of a mass and
a timescale then the first term naturally scaled directly as the mean
density and the latter term, the timescale, would only have to scale
as $\rho^{-1/2}$ in order to give $n = 1.5$, a value which seemed to be
broadly supported by the observations of the time. That is, the star
formation rate $f_{SFR}$ = $M_*/\tau_{SFR} \propto
\rho_g/{\rho_g}^{-1/2} \propto {\rho_g}^{3/2}$. Many gravitationally
controlled timescales depend on ${\rho_g}^{-1/2}$ so there was some
early, but still {\it a posteriori}, justification for this particular
interpretation.

However gravity is not the only mechanism operating in the production of
stars.  And so we ask what are the various modes of star formation
that might be considered in trying to predict an appropriate timescale
for star formation? And what might the controlling parameters be for
those different modes? Here we simply enumerate four main modes that
appear to be operating, and then we focus our attention primarily on
the first, quiescient star formation. Lessons learned here, we
believe, can be applied in provisionally parameterizing the other
modes of star formation as well.

Four distinct modes of star formation, whose rates are individually
determined by independent physics and very different timescales,
immediately suggest themselves: (1) a secular timescale $\tau_c$ for
unforced, that is to say, quiescent star formation across the galaxy
in question, regulated primarily by the natural cloud coalescence and
collapse timescales, (2) an induced star-formation timescale,
$\tau_d$, which is determined by quasi-periodic phenomena that are
internal to the galaxy, such as density waves, rotating bars, etc.,
(3) an impulsive star formation timescale, $\tau_e$, which is
determined by external, and generally aperiodic encounters or
collisions with other nearby galaxies or satellites, and finally (4) a
flow timescale, $\tau_f$, determining the rate of star formation
particularly in the nuclear regions of (typically starburst) galaxies,
where angular momentum loss mechanisms are probably the rate-limiting
factors.

\subsection{Quiescent Star Formation}

In this {\it Letter} we focus exclusively on parameterizing the
quiescent mode of star formation. Starting afresh, from dimensional
arguments alone we develop a simplified parameterization of the rate
of star formation that now captures the essential non-linear form
(Bigiel et al. 2008) of the current observational correlations. In the
life cycle of gas, from its diffuse state into collapsing clouds,
through the phase change called star formation and finally the cloud's
disruption and redispersal before it begins the cycle again, two
critical timescales are identified: (1) a cloud
coalescence/(re)collapse timescale $\tau_c$, characterizing the
systematic assembly and/or re-assembly of neutral hydrogen into a
molecular cloud and its subsequent development of a dense core,
ultimately leading to the observed star formation event, and then (2)
a star-formation-induced ``stagnation'' timescale, $\tau_s$,
determined by the presence of hot stars and their disruptive
interaction with the birth cloud.

Such a characterization involving two independent timescales is not
just convenient, it is methodologically required, because what is done
observationally amounts to selecting regions that have a given column
density of gas and then equating the relative {\it areal frequency} of
those regions with a relative {\it temporal frequency} (i.e., a
cumulative timescale or lifetime) for that density.  However that
cumulative timescale must itself be the sum of two independent
timescales, $\tau_s$ and $\tau_c$, only one of which, $\tau_c$, is
assumed here to be a function of the gas density; the other timescale,
$\tau_s$, being a function of stellar astrophysics and gas
heating. The argument is that empirically one needs to take
observations, which are per force averages over areas, and map them to
a theory that involves averages over time. Those regions of fixed
surface density that define the areas are either occupied by currently
observabale star formation tracers or they are not. The occupied time
is the interval over which the most recently formed stars (and stellar
physics) control the gas dynamics; the unoccupied time is primarily
the cloud assembly and collapse time, where gas density (and more
generally gas physics alone) is assumed to be the controling
factor. These are two physically distinct and separable processes that
have independent timescales; however, it is their sum that enters into
the mapping from the time domain to any given, instantaneously
observed, spatial coverage.

The stagnation time, $\tau_s$, is ultimately a measure of the
(wavelength-independent) timescale over which the general presence of
ionization fronts, winds, radiation pressure, SNe, and other
disrupting effects of the star-formation events influence the gaseous
environment, by stalling and otherwise preventing the onset of the
next cycle of star formation in that same region. As such $\tau_s$ it
is not a timescale to be equated with the lifetime of an individual O
star or a B star per se, but rather it is a timescale to be associated
with ``the disruptive presence of stars, and the star formation event
in general''.  It is associated with stellar lifetimes but it is a
single time period for a given region and it will not be a function of
wavelength ($H_\alpha$, vs FUV) or any given selected tracer (O stars
or B stars). The best that can be said is that it will be at least as
long as the longest-lived (high-mass) stars and those by-products
(i.e., Type II supernovae) that influence the structure of the local
medium in a way that they collectively delay the onset of the next
re-coalescence and collapse sequence. Some fraction of the stagnation
time will also be coupled to the time it takes gas to recombine after
being ionized by the HII region, or to cool back down after being
shocked by a supernova blast wave, and both of these are admittedly
strong ($n^2_e$) functions of electron density, but since the observed
control parameter being considered here is the neutral and/or
molecular gas density, to first order, $\tau_s$ should be decoupled.

\begin{figure}
\includegraphics [width=16cm, angle=0] {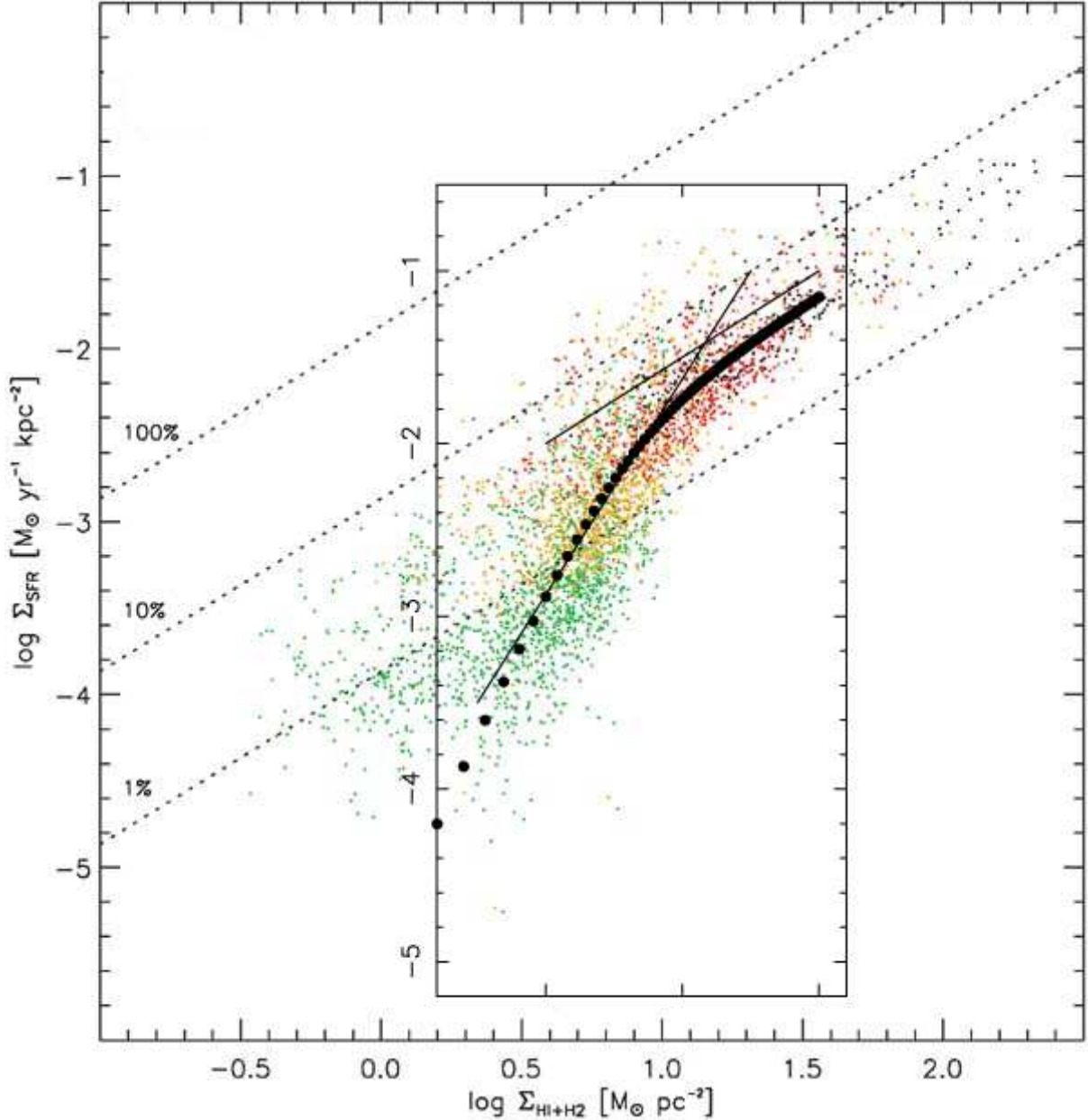}
\caption{\small The rate of star formation $\Sigma_{SFR}$ versus total gas
(molecular + neutral hydrogen) surface density $\Sigma_{H_2 + HI}$ for
sub-kiloparsec sampled galaxies in Bigiel et al (2008), with the
appropriately normalized prediction from Equation 1 superimposed on
his Figure 8. The asymptotic linear (unit slope) behavior predicted by
Equation 1 is seen at high gas surface densities with an increasingly
steeper fall-off toward low gas densities. This particular (discretely
sampled) solution is for illustrative purposes only; it is chosen as a
representative fit to the general trend in the cumulative
plot. However, it does show show how the linear regime is approached
at high gas surface densities, and in this particular example the
solution asymptotically approaches a power law with $n = 2.5$ at low
gas densities.  The dotted lines of unit slope crossing the diagram
correspond to the levels of constant star formation efficiency needed
to consume 1, 10 and 100\% of the gas reservoir in $10^8$ years. The
data points are color-coded by their respective radial distances
scaled by $r_{25}$ (black $<$ 0.25, red 0.25-0.50, orange 0.50-0.75,
and green $>$ 0.75). Detailed fits to individual galaxies will be
forthcoming.}
\end{figure}

We equate the ``star formation rate'' $f_{SFR}$, with the mass of
recently formed stars $M_*$ divided by some characteristic timescale
$\tau$ such that $f_{SFR} = M_*/\tau$. We now assume some efficiency
$\epsilon$ for converting into stars the total available gas mass
$M_g$, contained in some fixed volume $V$, such that $M_* = \epsilon
M_g = \epsilon V\rho_g$. Furthermore we break $\tau$ down into its two
rate-limiting timescales, $\tau_s$ and $\tau_c$, which, as mentioned
above, are the star-formation-induced, cloud-disruption  timescales and
the neutral-to-molecular cloud formation/collapse timescales,
respectively. Accordingly,

$$f_{SFR} = {M_* \over{\tau}} = {\epsilon V \rho_g \over{(\tau_s + \tau_c)}}$$

$$ ~~~~~~~  = {A \rho_g\over{(\tau_s + B (\rho_g/\rho_c)^{-n})}}$$

$$ ~~~~~~~  = {A\rho_g/\tau_s \over{(1 + (B\rho_c^n/\tau_s)\rho_g^{-n})}}$$

$$ ~~~~~~~  = {\alpha\rho_g \over{(1 + \beta\rho_g^{-n})}} \ \ \ \ \ \ \ \ (1)$$

\par\noindent where $A = \epsilon V$, $\alpha = A/ \tau_s$, $\beta =
(B\rho_c^n\tau_s)$ and it is assumed that $\tau_c = B(\rho_g/\rho_c)^{-n}$.

\section{Discussion and Conclusions}

The above derivation was made for quiescent star formation, wherein it
is implicitly assumed that the rate-limiting timescale behind this
particular mode is the coalescence and collapse time of the ambient
gas into molecular clouds, which subsequently act as the more
immediate sites for star formation. All of the sub-resolution physics
and astrophysics of cloud formation and collapse is secreted away in
this simple parameterization on the cloud stagnation timescales
($\tau_s$) and local gas densities ($\rho^{-n}$). Other modes of star
formation of course exist, and because of the very different physics
involved, they deserve their own parameterization; but the simple
timescale framework should in principle accommodate these other modes
as well. The collisionally driven (impulsive) star formation rate
$\tau_c$ would have a timescale set by the local galaxian environment.
The internally driven star formation rates due to density waves would
have their timescales $\tau_D$ strictly set locally at a given
galactocentric radius by 1/$\mid \Omega_{pattern} - \Omega_{rotation}
\mid$, where the denominator is the rate of passage of material
through the density wave of fixed pattern speed. Finally, one could
imagine finding an appropriate timescale $\tau_E$ for the transfer of
angular momentum regulating the flow of material into the central
regions of galaxies where starburst activity then occurs. In any case
the simple dimensionality of the proposed parameterization guarantees
some measure of success; the novelty is in decoupling and making
explicit the star formation tracer's visibility timescale from the
independent assembly/formation/collision timescale.  

There are, of course, other proposed parameterizations of star
formation rates, many of which are far more bottom-up and physically
based rather than the top-down and more phenomenologically motivated
approach historically made and adopted here.  Dopita has long argued
for a ``compound'' Schmidt Law (e.g. Dopita \& Ryder 1994 and
references therein) that depends both on local gas density and the
total surface density of matter.  Silk (1997) argues that the SFR is
self-regulated with gravitational instability playing off against
supernova heating of the interstellar medium. Tan (2000) suggested
that the cloud-cloud collision timescale was the rate-limiting factor
for star formation; while Krumhotz \& McKee (2005) suggested a simple
linear relation holds between the surface density of molecular
hydrogen and SFR based on the suggestion that the free-fall time for
giant molecular clouds is a largely independent of the cloud mass
(Solomon et al. 1987). Li, Mac Low \& Klessen (2005) investigated the
role played gravitational instabilities controling star formation
rates using a three-dimensional, smooth particle hydro
code. Simulations by Dobbs \& Pringle (2009) simply tie the SFR to the
local dynamical timescale of the gas.  Finally, Blitz \& Rosolowsky
(2006 and earlier references therein) suggest that mid-plane pressure
is the controling factor in converting HI into its molecular phase and
then on to star formation. A detailed comparison of the predictions of
some of these models, as well as a number of thresholding scenarios,
with the available observations has been given in Leroy et al. (2008)
and the interested reader is referred to that paper for a detailed
discussion of their relative merits and degrees of success in matching
the observations.

The observed rate of star formation is further complicated in the high
density nuclear regions of galaxies that are probably not undergoing
typical, quiescent star formation that is the main focus of this {\it
Letter}. At very high column densities, often only found in the
nuclear regions of galaxies, the raw correlation of star formation
rates with very high gas column densities appears to steepen with
respect to the relation found the lower-density main disk. Assuming
that the molecular fraction has been properly estimated and/or
measured, we have two possible explanations for these observations,
both of which may be operating in any given situation. First, the
plane thinkness in these inner regions may be playing a factor in
increasing the volume gas density without necessarily changing the
apparent column density. Adopting a standard (outer disk) Schmidt law
would result in the observed upturn in the correlation because of the
increased volume gas density and the non-linear response in the
production of stars each of which are only measured in projection.

In addition to, or independently of, any plane thickness compression
at very high (nuclear) gas densities the same steepening of the
observed relation can be easily induced by another plausible effect,
beam dilution. If, within the naturally imposed or artificially
selected resolution element empirically used to make the star-tracer
and gas density correlations, the actual size of the region undergoing
star formation is smaller than the beam then, as in the case of the
decreasing plane thickness situation above, the appropriate volume
density of the region actually undergoing star formation will be much
higher in reality than the beam-smeared and diluted column density
would suggest.  If this is indeed the case, then the observed
correlation of star density versus projected gas surface density would
steepen, because of the highly non-linear underlying sensitivity of
the rate of star formation to volume density.

\medskip
\medskip
\acknowledgments

This study made use of the NASA/IPAC Extragalactic Database (NED)
which is operated by the Jet Propulsion Laboratory, California
Institute of Technology, under contract with the National Aeronautics
and Space Administration. My thanks to Kay Baker for her help with the
figures. Thanks also to Samuel Boissier, Bruce Elmegreen, T.J Cox and
Wendy Freedman for comments on early drafts of this paper. And
finally, my sincere thanks to the anonymous referee who gave the paper
a very careful reading and prevailed upon me to clarify why yet
another parameterization was needed in light of the many competing
recipes already available in the literature.

\medskip
\medskip
\vfill\eject
\medskip
\medskip

Blitz, L. \& Rosolowsky, E. 2006, \apj, 650, 933

Bigiel, F., Leroy, A., Walter, F., Brinks, E., de Blok, W.J.G., Madore, B.F., \& Thornley, M.D.  2008,\aj, 136, 2846

Dobbs, C.L., \& Pringle, J.E. 2009, \mnras, 396, 1579

Dopita, M.A., \& Ryder, S.D. 1994, \apj, 430, 163

Hamajima, K., \& Tosa, M. 1975, \pasj, 27, 561

Hartwick, F. D. A. 1971, \apj, 163, 431

Krumholz, M.R., \& McKee, C.F.J. 2005, \apj, 630, 250

Li,Y., Mac Low, M.M., \& Klessen, R.S.  2005, \apj,
620, L19

Leroy, A., Walter, F., Brinks, E., Bigiel, F., de Blok,
W.J.G., Madore, B.F., \& Thornley, M.D.  2008, \aj, 136, 2782

Madore, B.F. 1977, \mnras, 178, 1

Madore, B.F. van den Bergh, S., \&  Rogstad, D.H. 1974, \apj, 191, 317

Nakai, N., \& Sofue, Y. 1984, \pasj, 36, 313

Sanduleak, N. 1969, \aj, 74, 475

Schmidt, M. 1959, \apj, 129, 243

Schmidt, M. 1963, \apj, 137, 758

Silk, J. 1997, \apj, 481, 703

Tan, J. C. 2000, \apj, 536, 173

Tosa, M., \& Hamajima, K. 1975, \pasj, 27, 501

\vfill\eject
\end{document}